\documentclass[aps,pre,twocolumn,float]{revtex4}
\usepackage{amsmath,bm,epsfig}
\usepackage{amssymb}
\newcommand{\beq}{\begin{equation}}
\newcommand{\eeq}[1]{\label{#1}\end{equation}}

\begin{document}
\title{Elasticity in Amorphous Solids: Nonlinear or Piece-Wise Linear?}
\author{Awadhesh K. Dubey, Itamar Procaccia, Carmel A.B.Z. Shor  and Murari Singh}
\affiliation{Dept. of Chemical Physics, the Weizmann Institute of Science,  Rehovot 76100, Israel}
\begin{abstract}
Quasi-static strain-controlled measurements of stress vs strain curves in macroscopic amorphous solids result in a
nonlinear looking curve that ends up either in mechanical collapse or in a steady-state with fluctuations
around a mean stress that remains constant with increasing strain. It is therefore very tempting to fit
a nonlinear expansion of the stress in powers of the strain. We argue here that at low temperatures
the meaning of such an expansion needs to be reconsidered. We point out the enormous difference between quenched and
annealed averages of the stress vs. strain curves, and propose that a useful description of the mechanical response is given by a stress (or strain) dependent shear modulus for which a theoretical evaluation exists. The elastic response is piece-wise linear rather than nonlinear.
\end{abstract}
Version of \today
\maketitle
\section{Introduction}
\label{intro}
Materials designated as ``amorphous solids" span a large class of non-crystalline materials that exhibit
an elastic response to small strains or stresses. In this class one finds ``tough" materials like metallic
glasses as well as ``soft" materials like foams, with many intermediate (in strength) materials in between.
All this host of materials display initially a linear response to a quasi-static external loading (strain $\gamma_{ij}$  or stress $\sigma_{ij}$) with a shear
modulus that relates the stress to the strain. Omitting tensor indices for notational simplicity one writes
\begin{equation}
\sigma = \mu \gamma\ , \quad \gamma\ll 1 \ ,
\end{equation}
with $\mu$ being the shear modulus.

\begin{figure}
\includegraphics[scale = 0.60]{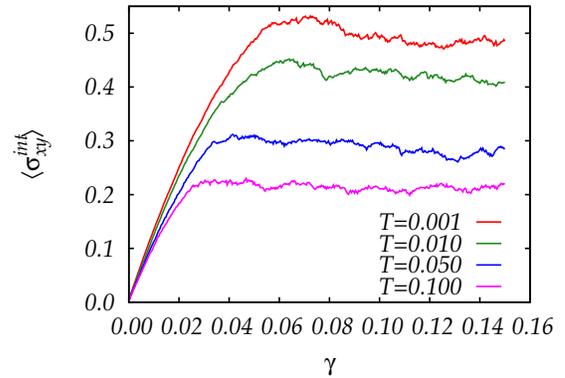}
\caption{Typical stress vs. strain curves obtained from molecular simulations using a 2-dimensional Kob-Andsersen
glass former with 65:35 ratio of 1000 point particles having Lennard Jones interaction with longer and shorter
interaction lengths. The curves shown are obtained by averaging 100 individual stress vs. strain curves obtained from 100 realization of the initially prepared glass. The temperatures shown are in Lennard-Jones units as explained in the text. Customarily one fits a nonlinear expansion like Eq.~(\ref{nonli}) to such curves. We argue in this paper that such nonlinear
expansions are not tenable.}
\label{averages}
\end{figure}

Upon the increase in the external loading this linear relation appears to fail. The response of the amorphous solid
begins to mix elastic intervals interspersed with plastic events \cite{99ML,03RR,04ML,10KLP,11ML,15LGRW}, leading generically to an apparent nonlinear
dependence of the stress as a function of the strain, see Fig.~\ref{averages} as an example.  The stress vs.
strain curves for large values of the strain either end abruptly due to a catastrophic failure of the material
or display a regime of ``steady state" where the shear modulus $\mu$ appears to vanish. Viewing
stress vs. strain curves of this type one is tempted to present them before the onset of the steady state as a nonlinear expansion, referred to as ``nonlinear elasticity", in the form (again with tensor indices omitted)
\begin{equation}
\sigma = \mu(\gamma=0) \gamma + B_2(\gamma=0)\gamma^2 +B_3(\gamma=0)\gamma^3+\dots \ . \label{nonli}
\end{equation}
The aim of this Letter is to discuss the validity of such expansions for generic amorphous solids at low temperatures, and in fact to argue that they should be carefully reconsidered. The discussion will also make clear what do we mean by ``low temperatures".

It is already known that at zero temperature $T=0$ the nonlinear coefficients in Eq. (\ref{nonli}) do not
exist in the thermodynamic limit \cite{11HKLP}. The issue raised here is what is the nature of the elastic
response of amorphous solids at finite temperatures. We will show that in fact one should consider the elastic
response of amorphous solids between plastic events, and there one can invariably define a ``piece-wise linear"
elastic response in the form
\begin{equation}
\Delta \sigma = \mu(\gamma) \Delta \gamma\ , \quad \text{for any value of $\sigma$ and $\gamma$} \ , \label{muofgamma}
\end{equation}
with $\mu(\gamma)$ determined theoretically, cf. Eq.~(\ref{mutheoretical}) below.
We should stress that this piece-wise linear law is also valid in the  steady
state regime with a finite shear modulus in spite of the apparent flat dependence of the stress on the strain. The
difference between the approach of Eq.~(\ref{nonli}) and the proposition Eq.~(\ref{muofgamma}) requires a discussion
of the difference between quenched and annealed averages.

To establish the proposed ``law" Eq.~(\ref{muofgamma}) one needs to examine first small systems at sufficiently
small temperatures (to be determined below) and
learn how to approach the thermodynamic limit. In this paper we will use quasi-static strain controlled protocols
with simple shear. The advantage of using
small systems and low temperatures is that one can resolve the stretches of strain for which the response of the system is
purely elastic as these are punctuated by intervening plastic events. The discussion below will be exemplified by
molecular dynamics simulations at low temperatures. Obviously
in such simulations one deals with systems that are not in thermodynamic equilibrium whose attainment requires
astronomical relaxation times. Nevertheless we will show that by increasing the strain quasistatically we can
equilibrate the systems in the restricted sense that between plastic events the average stress and the average energy reach stationary values. We refer to such states as ``restricted temperature ensembles";
while not in true thermal equilibrium they nevertheless succumb to thermal statistical mechanics.
In other words we will demonstrate that for our glasses the shear modulus
can be computed for any value of the strain by using the thermal expression \cite{89Lut,69SHH,13WXPWB}
\begin{equation}
\mu(\gamma) = \mu_{\rm B}(\gamma) -\frac{V}{k_BT} \left[\langle \sigma^2\rangle -\langle \sigma\rangle^2\right]  .
\label{mutheoretical}
\end{equation}
where $\mu_{\rm B}$ is the usual \cite{Born,12CP} Born approximation for the shear modulus and $V$ and $T$ are the volume and the actual temperature of the glass. $k_B$ is Boltzmann's constant. The stress fluctuations are measured as usual in an equilibrated Gibbs ensemble. Below we will use notation
\begin{equation}
\mu_F (\gamma) \equiv  \frac{V}{k_BT} \left[\langle \sigma^2\rangle -\langle \sigma\rangle^2\right] \ .
\end{equation}
\section{Numerical Simulations}

To generate data for the present discussion we perform molecular dynamics simulations using the Kob-Andersen
model, in which point particles interact via a Lennard-Jones potential. There are two types of particles A
and B and the parameters of the interaction potentials can be found in Ref.~\cite{14DIMP}. The system is prepared
by firstly
randomizing the particles in a volume V, and then we run molecular dynamics at $T=0.8$ in Lennard-Jones units
for which the Boltzmann constant $k_{\rm B}=1$. After equilibration the system is quenched, again using
molecular dynamics, to temperature $T=0.001$ at a rate of $\dot T=10^{-6}$. Simulations are performed at this
final temperature until the mean energy of the system stabilizes, exhibiting a time independent value. Simulation of
quasi-static straining are then performed either at this temperature or at any desired higher temperature
which is obtained by heating up the system up. After reaching the desired temperature one waits again
for the stabilization of the mean energy. Of course the time taken for stabilization are much shorter than
the glass relaxation time (known as $\tau_\alpha$) and the systems considered are {\em not} in true
thermal equilibrium. We will argue however that they reach a ``restricted" Gibbs ensemble that allows for
the definition of meaningful statistical averages.

\begin{figure}
\hskip -0.5 cm
\includegraphics[scale = 0.37]{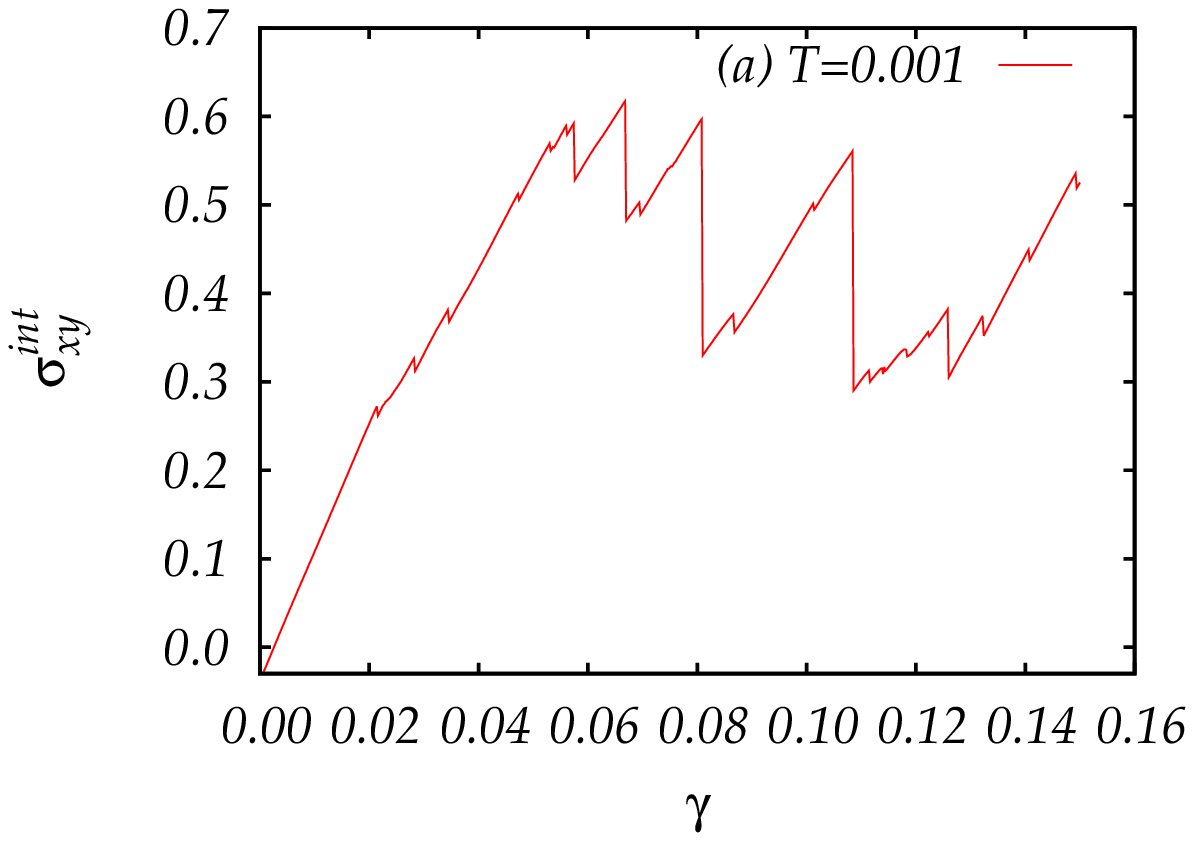}
\hskip -0.4 cm
\includegraphics[scale = 0.37]{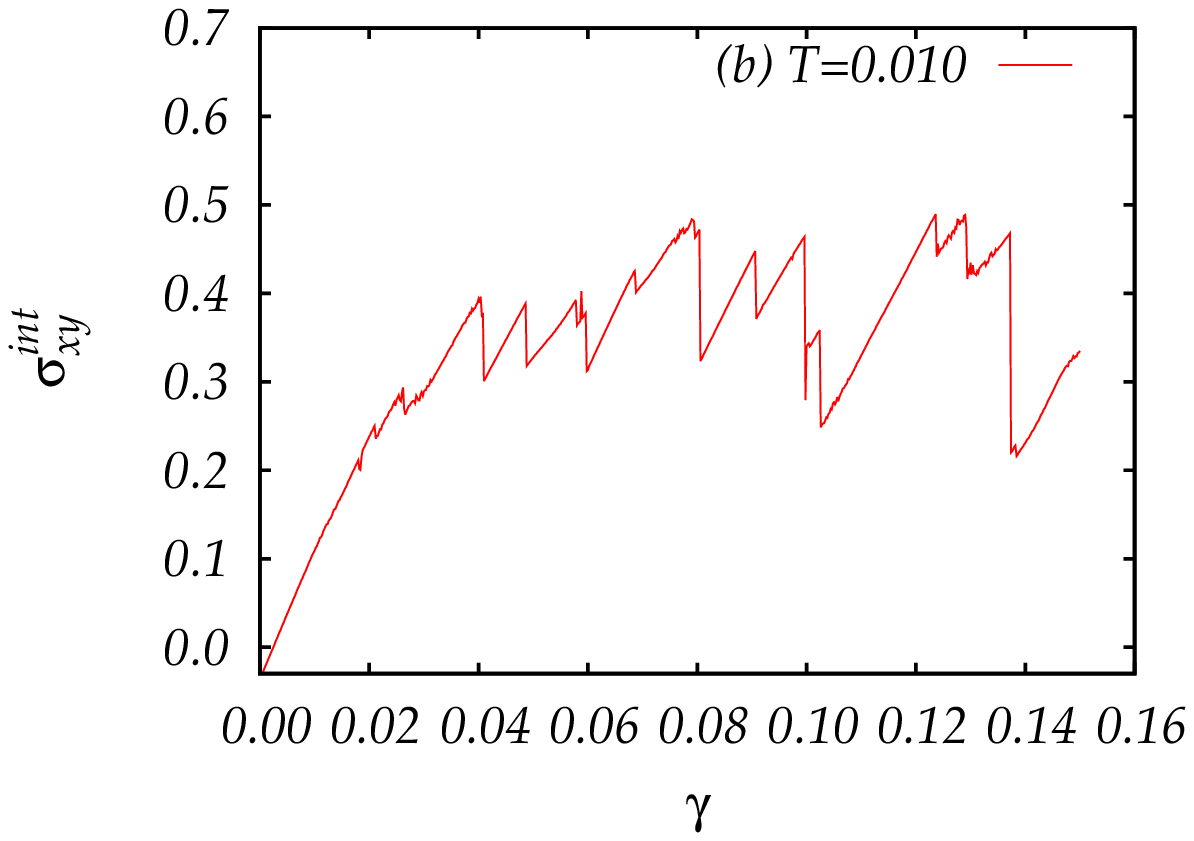}\\
\hskip -0.5 cm
\includegraphics[scale = 0.37]{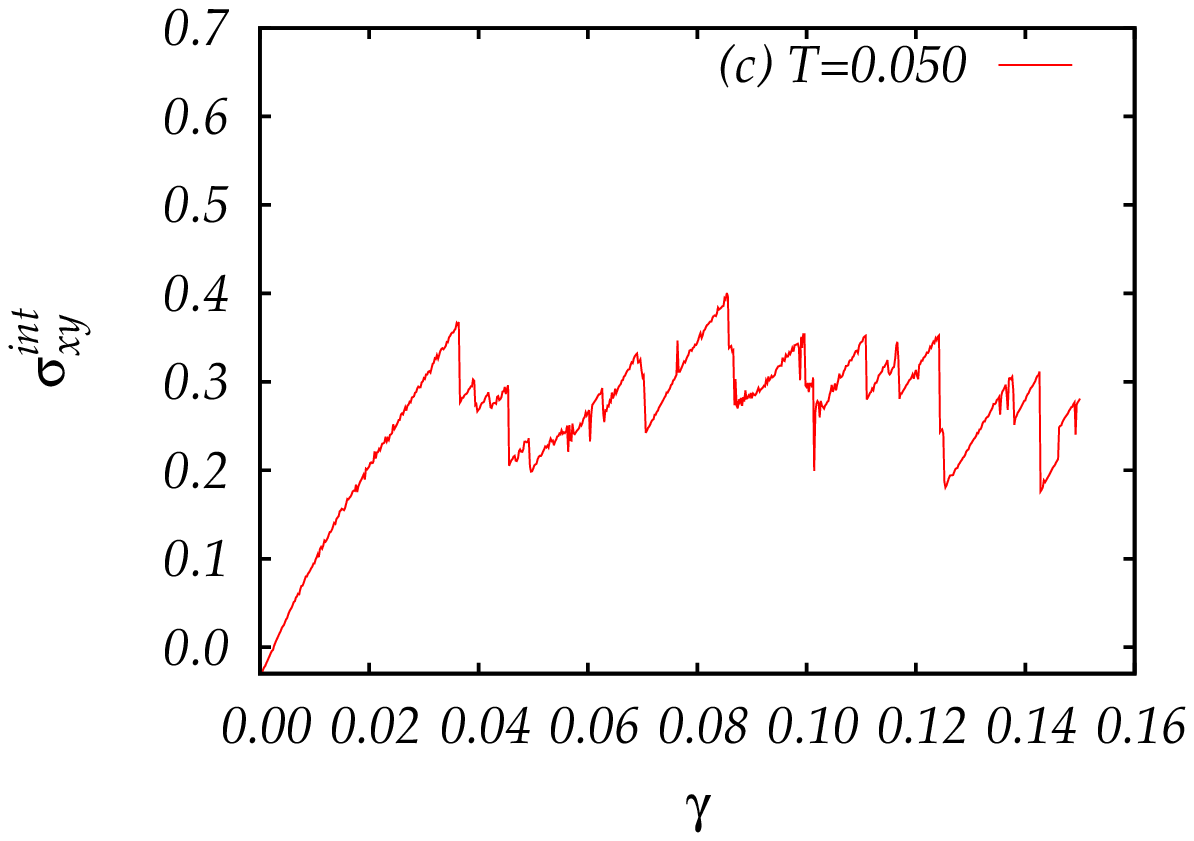}
\hskip -0.4 cm
\includegraphics[scale = 0.37]{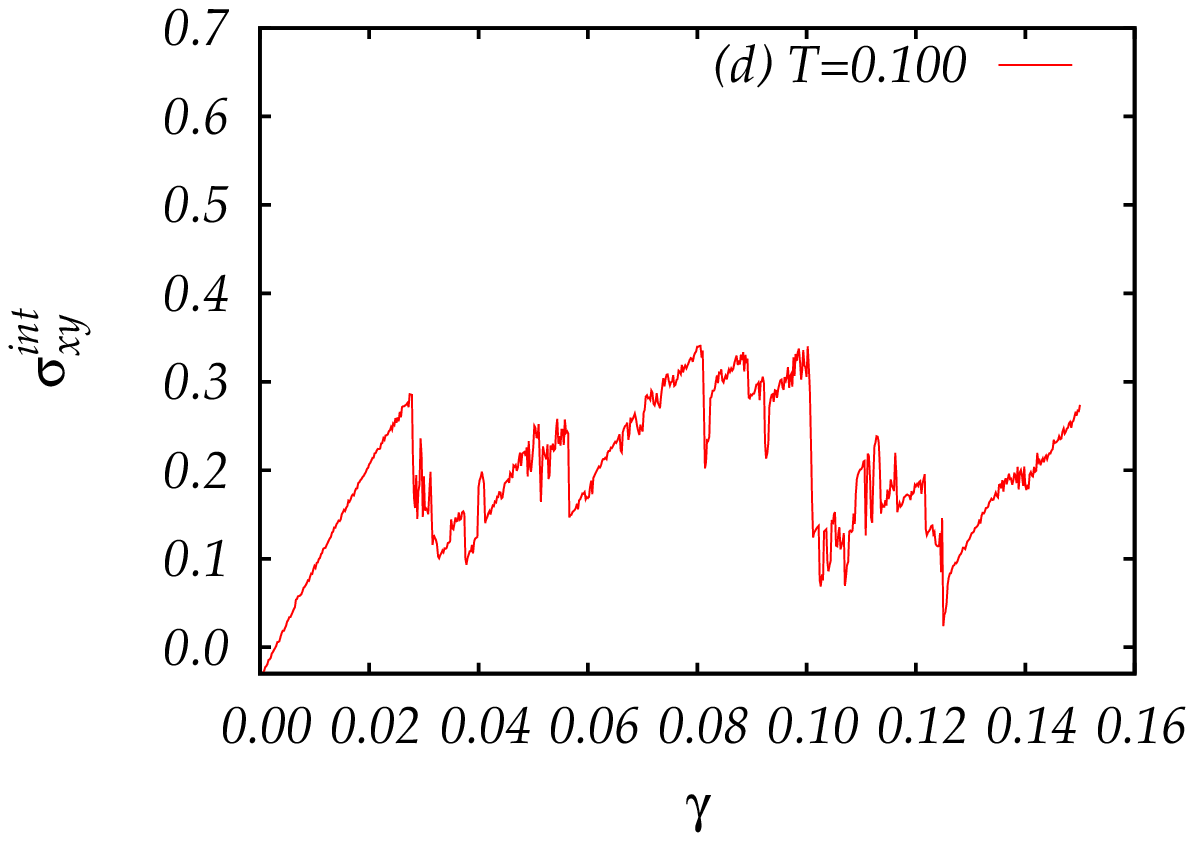}
\caption{Individual realizations of stress vs. strain curves for systems of 1000 particles at different temperatures.
Note that individual plots never attain a zero shear modulus between plastic events.}
\label{realizations}
\end{figure}

Once the system has stabilized its mean energy we strain it quasi-statically using simple shear
with Lees-Edwards periodic boundary conditions\cite{72LE}. Strain steps of magnitude $\delta \gamma=2\times 10^{-4}$ are taken, allowing the system to stabilize both its mean energy and mean stress for 500 MD steps after every such increase. Stabilization is then obtained when the average energy and the average stress are constant over additional 1000 MD steps. Next the mean stress $\sigma_{xy}^{int}$ and the second moment of the stress $\overline{\sigma^2}$ are measured for each realization. At this point we record the mean stress as a function of the strain. Finally we average the first and second moments of the stress
over our different realizations to obtain $\langle \sigma \rangle$ and $\langle \sigma^2\rangle$. With systems of 1000 particles the stress vs strain curves
for every realization reveal different characteristics from the averaged curves shown in Fig.~\ref{averages}. Typical such
stress vs. strain curves for individual realizations are shown in Fig.~\ref{realizations} at different temperatures.
The different realization show clearly the piece wise linear stretches between plastic event in which the stress drops suddenly. The nature of the plastic events changes before and after the ``yield" which is followed by the steady state regime.
Before yield the events are small, and they are known at $T=0$ to be represented by individual Eshelby quadrupolar displacement field that are associated with small energy drops that are system-size independent. Beyond yield at $T=0$ the events are system spanning events in which the energy drops are sub-extensive \cite{10KLP}. Also at finite temperatures one sees in Fig.~\ref{realizations} the change from small stress drops to large ones \cite{10KLPZ}. The highly serrated nature of the stress vs. strain curves is masked by the averaging of many realizations shown in Fig.~\ref{averages}, as well as in the thermodynamic limit. For large systems the density of events increases enormously as the strain intervals between plastic events decrease rapidly with the system size \cite{15HJPS}. Accordingly, we need to examine carefully the elastic response of the system and how to interpret the thermodynamic limit in a meaningful way.
\begin{figure}
\hskip -0.5 cm
\includegraphics[scale = 0.37]{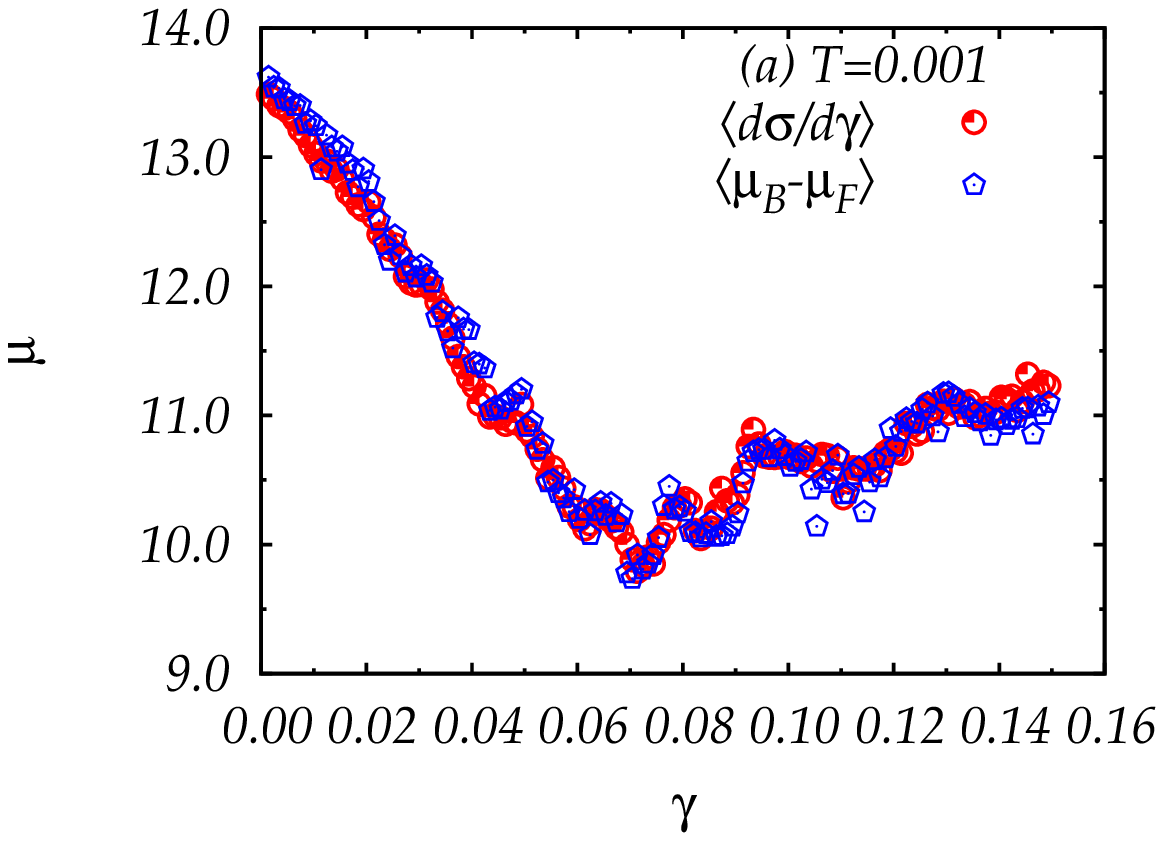}
\hskip -0.4 cm
\includegraphics[scale = 0.37]{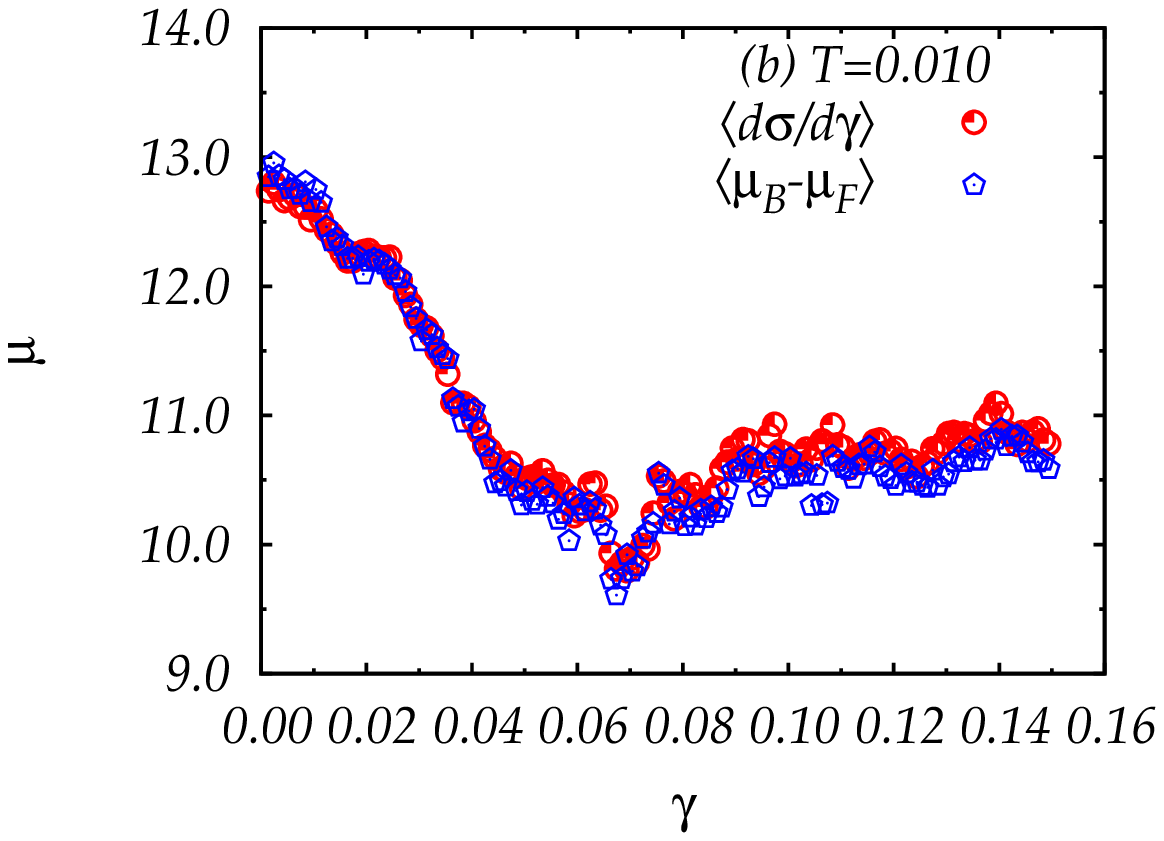}\\
\hskip -0.5 cm
\includegraphics[scale = 0.37]{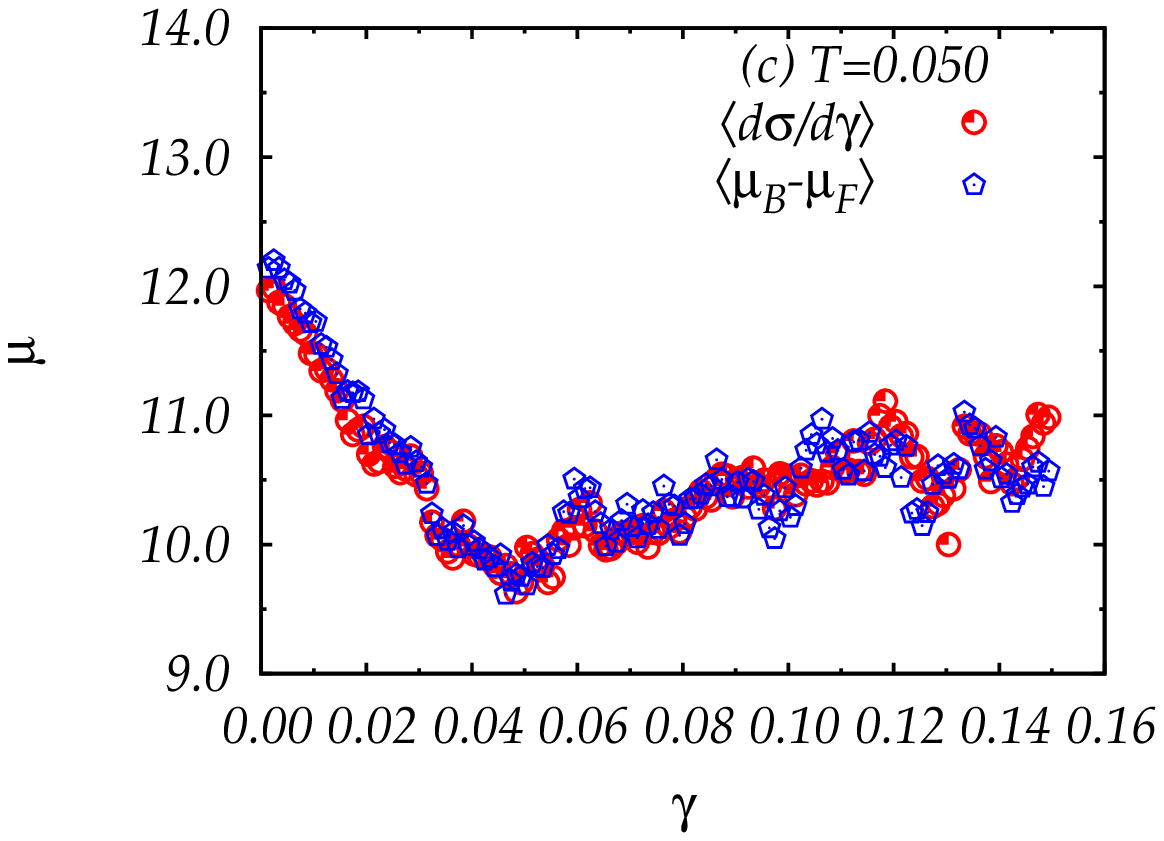}
\hskip -0.4 cm
\includegraphics[scale = 0.37]{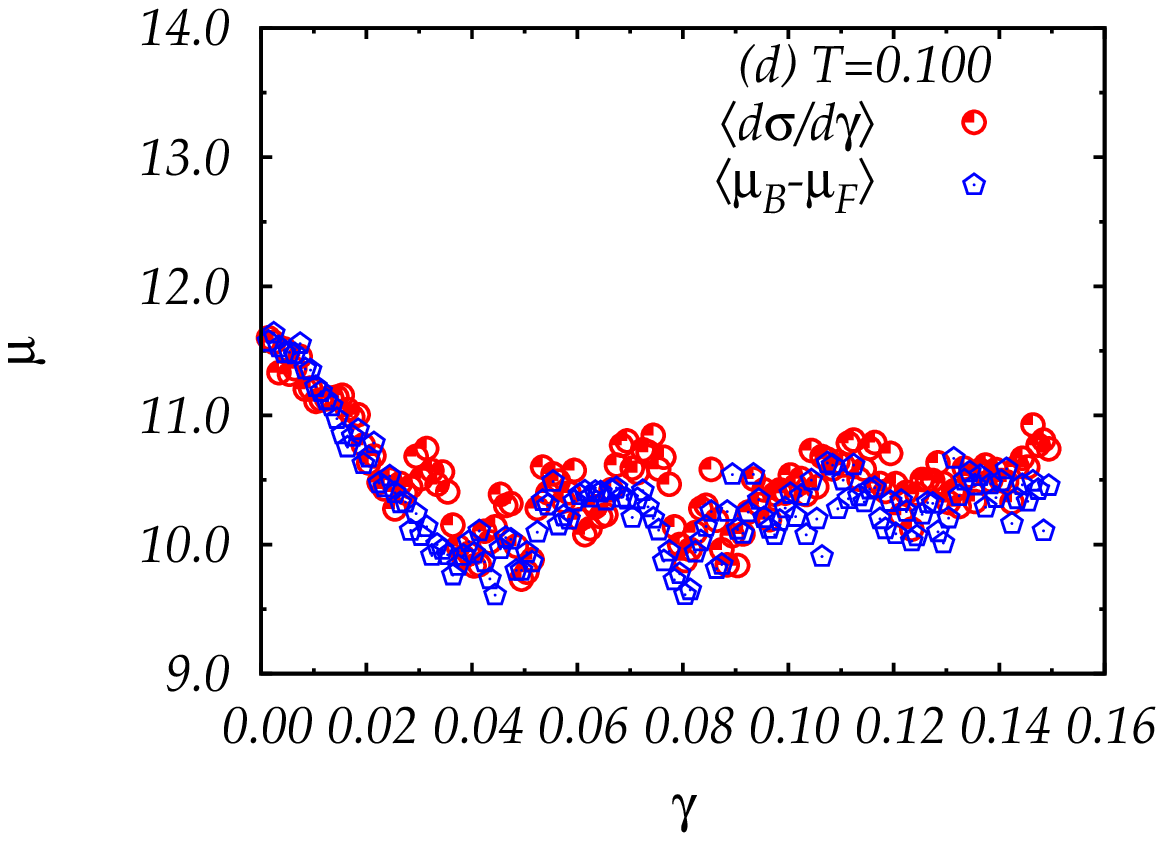}
\caption{Comparison of the direct measurement of quenched averages of the shear modulus from the local slope of the
strain vs. stress curves and the theoretical expression Eq.~(\ref{mutheoretical}).}
\label{compare}
\end{figure}
\section{The Elastic Response and quenched averages}
Denote the sequence of strain increments as $\Delta \gamma_i$ and the thermally averaged stress measured after
the system stabilizes as $\sigma_i$. To calculate the shear modulus $\mu$ from the slope of the thermally averaged stress vs strain we determine a sequence of values $\sigma_i$ that satisfies $\sigma_{i+1}>\sigma_i$ and compute the average slope
where now the average is over the sequence. Once we have found a couple of stress values satisfying $\sigma_{i+1}<\sigma_i$
we begin a new sequence of increasing stress values and average again the slope for that sequence. We associate
the value of this average slope with the mean strain in the sequence.
At this point we introduce the quenched averages. Measuring then the observed value of the slope we
average it over O(100) realizations at the same values of $T$ and in a small bin of $\gamma$ values, and denote the results as $\mu(\gamma, T)$.
In parallel, for each measured thermally averaged stress we compute also the stress fluctuations, and evaluate the Born approximation directly from the known
Hamiltonian of the system. We then compute the expression for the shear modulus as provided by Eq.~(\ref{mutheoretical}).
As before, we average the resulting number over O(100) realizations. The two evaluations of the shear modulus are compared
in Fig.~\ref{compare}.

This comparison allows us to reach a number of important conclusions: (i){\em The two evaluations agree}. This means that the actual elastic response should be considered piece-wise linear rather than resulting from a nonlinear expansion. In no way can one say that the local slope is, say, given by $\mu(\gamma=0) + 2 B_2 \gamma +3B_3\gamma^2+\dots$. Throughout the strain range the stress fluctuations reduce the Born term to predict correctly the local linear response of the system. This includes
the steady state regime of the strain controlled protocol where it is customary to take the shear modulus as zero. (ii) For the higher temperatures the evaluation of the local slopes should be done with care, since temperature fluctuations begin to introduce
``spurious" apparent slopes in the stress vs strain curves.  The temperature $T=0.1$ is the highest temperature for this particular system
for which we can trust the procedure. At higher temperatures there are too many intervals with $\sigma_{i+1}<\sigma_i$; it becomes too difficult to separate mechanical increases
of stress due to strain changes from random temperature fluctuations in the stress. In this sense we are limited
to ``sufficiently low temperatures".
\section{Annealed averages}
\begin{figure}
\hskip -0.5 cm
\includegraphics[scale = 0.50]{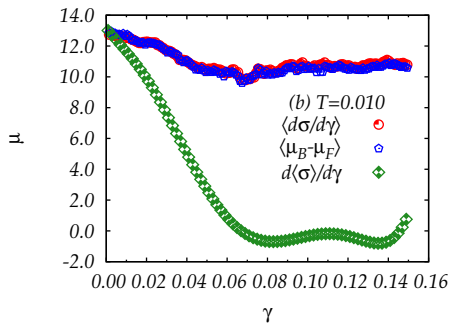}
\caption{An example of the comparison of annealed and quenched averages of the local slope of the
strain vs. stress curves and the theoretical expression Eq.~(\ref{mutheoretical}). This example is at $T=0.01$ but
the conclusion is identical in all the tested temperatures: except at
$\gamma=0$ the results of the annealed and the quenched averages of the stress vs. strain curves
differ greatly.}
\label{nogood}
\end{figure}

\begin{figure}
\includegraphics[scale = 0.30]{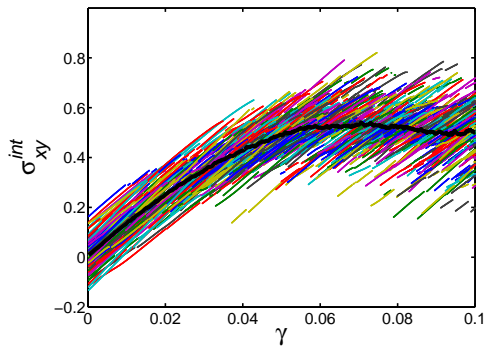}
\caption{The annealed stress vs. strained curve (in continuous black line) and the actual
segments of linear response of the various realizations that were annealed to get the the continuous line.
Except at $\gamma=0$ the annealed procedure does not supply the right information regarding the mechanical response.}
\label{carmel}
\end{figure}

At this point we need to discuss the meaning of the local slope of the annealed average of the stress vs. strain curves
that are shown for example in Fig.~\ref{averages}. We already mentioned that nonlinear expansions in $\gamma$ around
$\gamma=0$ are untenable at $T=0$, and probably also at finite but low temperatures. The question is whether the
local slope of such curves yields a number that is the same, or close to, our $\mu(\gamma)$. The answer is negative as
can seen in Fig.~\ref{nogood}.
It turns out that averaging the stress vs. strain curves before computing the local slopes results in
a smoother looking curve which is nevertheless not really differentiable. It is made of individual contributions in
which every plastic drop is a singular event where differentiability is lost. Since these events occur at different
values of $\gamma$ for each realization, averaging results in smoother looking curves, but the local slope
of the resulting annealed average is {\em not} the shear modulus as computed from its theoretical definition Eq.~(\ref{mutheoretical}). To make this point crystal clear we show in Fig.~\ref{carmel} the annealed averaged
stress vs. strain curve at $T=0.001$ and superpose on it the actual realizations that give rise to it by
averaging. As said, the discrepancy is most glaring in the steady state regime, but it is equally severe
at each value of $\gamma\ne 0$.

Finally, we need to discuss the relevance of these findings to the macroscopic experiments in which
it is impossible to resolve the tiny stress drops which are extremely dense, resulting in a relatively
smooth looking curve. If we accept the tentative view that such curves are equivalent to our annealed
procedure due to self-averaging, then we must conclude that their local slope is not providing a correct
measurement of the shear modulus as a function of $\gamma$. The discrepancy is of course most glaring in
the steady state regime, where the annealed protocol results in a vanishing shear modulus. This is
clearly incorrect since the strained controlled system can support a stress without flowing, meaning that the
shear modulus cannot be zero. Here we find the the discrepancy occurs already in the so-called ``elastic
regime". To get a correct measurement of $\mu(\gamma)$ one
should measure the stress fluctuations and estimate the Born term to compute Eq.~(\ref{mutheoretical}). This is
of course not an easy task, but the discussion presented above should provide a warning as to the proper
interpretation of the local slopes of macroscopic stress vs. stress curves.

\section{Concluding remarks}

The main point of this paper is that in random systems one can expect that quenched and annealed averages might
yield different results. This is shown to be particularly true for the shear modulus, which can be computed
either for each realization and then averaged, or rather from an average of stress vs strain curves. The answer
is very different as can be seen from Fig.~\ref{nogood}. The question ``which is then the relevant shear modulus" is
answered in our opinion by comparing with the theoretical expectation Eq.~(\ref{mutheoretical}). This theoretical
expression is derived for system in thermal equilibrium. Another result of the present study is that although
quenched glasses are not at true thermal equilibrium, they can be thermalized in the restricted sense that their
average energy and stress are stationary. In that situation the stress fluctuations can be measured and the theoretical
value of the shear modulus is in very good agreement with the quenched rather than the annealed average as described
above. We therefore propose that measurements of shear moduli and other mechanical indices that employ annealed
or self-averages must be considered with extra care. These may not be the actual indices that are better revealed
by either quenched averages or by expressions of the type of Eq.~(\ref{mutheoretical}).

\acknowledgments
This work had been supported in part by an ERC ``ideas" grant STANPAS
and by the Minerva Foundation, Munich Germany.



\begin{thebibliography}{99}

\bibitem{99ML}
D. L. Malandro and D. J. Lacks, J. Chem. Phys. {\bf 110}, 4593 (1999).

\bibitem{03RR}
J. Rottler and M. O. Robbins, Phys. Rev. E {\bf 68}, 011507 (2003).

\bibitem{04ML}
C. Maloney and A. Lemaitre, Phys. Rev. Lett. {\bf 93}, 195501 (2004).

\bibitem{10KLP}
S. Karmakar, E. Lerner, and I. Procaccia, Phys. Rev. E {\bf 82}, 055103(R) (2010).

\bibitem{11ML}
M. L. Manning and A. J. Liu, Phys. Rev. Lett. {\bf 107}, 108302 (2011).

\bibitem{15LGRW}
J. Lin, T. Gueudr\'e, A. Rosso, and M. Wyart, Phys. Rev. Lett. {\bf 115}, 168001 (2015).

\bibitem{11HKLP}
H.G.E. Hentschel, S. Karmakar, E. Lerner and I. Procaccia, Phys. Rev. E {\bf 83}, 061101 (2011).
\bibitem{89Lut}
J.F.Lutsko, J. Appl. Phys. {\bf 65}, 2991 (1989).

\bibitem{69SHH}
 D. R. Squire, A. C. Holt, and W. G. Hoover, Physica {\bf 42}, 388 (1969).
\bibitem{13WXPWB}
J. P. Wittmer, H. Xu, P. Polińska, F. Weysser, and J. Baschnagel, J. Chem. Phys. {\bf 138}, 12A533 (2013).
\bibitem{Born}
 M. Born and H. Huang,
{\em Dynamical Theory of Crystal Lattices}
 (Oxford University Press, 1954) \ .

 \bibitem{12CP}
 Y.Cohen and I. Procaccia, Europhys. Lett. {\bf 99}, 46002 (2012).

\bibitem{14DIMP}
V. Dalidonis, V. Ilyin, P. Mishra and I. Procaccia, Phys. Rev. E, {\bf 90}, 052402 (2014).
\bibitem{72LE}
A.W. Lees and S.F. Edwards J. Phys. C: Solid State Phys. {\bf 5}, 1921 (1972).
\bibitem{10KLPZ}
S. Karmakar, E. Lerner, I. Procaccia, and J. Zylberg, Phys. Rev. E {\bf 82}, 031301 (2010).
\bibitem{15HJPS}
H.G.E. Hentschel, P. K. Jaiswal, I. Procaccia and S. Sastry, Stochastic Approach to Plasticity and Yield in Amorphous Solids, Phys. Rev. B., in press.





\end{thebibliography}
\end{document}